\documentclass[aps,pre,twocolumn,amssymb,amsmath,floatfix]{revtex4}
\usepackage{graphicx,subfigure}
\usepackage{bm}
\usepackage{verbatim}
\usepackage{amsmath}
\usepackage{amssymb}
\usepackage[T1]{fontenc}
\usepackage{ae,aecompl}

\newcommand{\h}{{\tilde h}}

\newcommand{\eps}{{\varepsilon}}

\newcommand{\oh}{{\frac{1}{2}}}

\newcommand{\cH}{{\mathcal H}}
\def\rf#1{(\ref{#1})}
\def\rfs#1{Eq.~\rf{#1}}

\begin{document}

\title{A note on a relation between ac Josephson effect and double-well BEC oscillations} 
\author{Leo Radzihovsky}
\author{Victor Gurarie}
\affiliation{Department of Physics, University of Colorado, Boulder, CO 80309}
\date{\today}

\begin{abstract}
  In this brief note we comment on the relation between the ac
  Josephson effect and the coherent oscillations of a Bose-Einstein
  condensate confined to a double-well potential. The goal is to
  elucidate the extent to which the latter is a realization of the
  former. We detail the correspondence that emerges in the high
  occupation limit of the double-well potential, and particularly note
  the relation between the two oscillation frequencies.
\end{abstract}

\maketitle

A realization of trapped degenerate atomic gases has opened
opportunities to study many interesting quantum many-body phenomena in
previously unexplored regimes\cite{DalfovoRMP1999}.  Recent
experiments\cite{AlbiezJEprl2005,SteinhauerJEnature2007} on
Bose-Einstein condensates (BEC), trapped and oscillating in an
imbalanced double-well potential have sought to realize an atomic Bose
gas analog\cite{JavanainenJEprl1986,SmerziJEprl1997,LeggetJEpra1998,
  GiovanazziJEprl2000,MeierJEpra2001} of the alternating- and
direct-current (ac and dc) Josephson
effects\cite{BrianJosephson,applJE}.

While there are some obvious analogies, considerable fundamental
differences between the two systems exist. For instance, a
conventional Josephson junction (JJ) between two superconductors is an
open quantum many-body system driven by a fixed electro-chemical
potential (voltage) difference between the left and right contacts, or
with a current imposed by an electrical circuit. Thus on general
grounds (gauge invariance and Heisenberg equation of motion) the
evolution of the relative phase $\phi=\phi_L-\phi_R$ in an ac
Josephson effect is given, {\em exactly}, by
\begin{eqnarray}
\hbar\dot\phi &=& 2e V,
\end{eqnarray}
where $e V=\mu_L-\mu_R$ is the imposed electro-chemical potential
difference across the left and right contacts controlled by the
voltage $V$. When combined with the (lowest harmonic) expression for
the Josephson current, $I=I_0\sin\phi$, the above exact {\em linear}
growth of $\phi(t)$ with time gives the standard ac Josephson effect,
\begin{eqnarray} \label{eq:vf} 
I(t)=I_0\sin(\omega_J t), \ \omega_J =
  2e V/\hbar.
\end{eqnarray}
In addition to its basic physical importance it provides an extremely
accurate voltage-frequency relation (thereby defining a Volt to one
part in $10^8$) with a number of other important
applications\cite{applJE}. In a conventional superconducting JJ the
critical current is given by the standard Ambegaokar-Baratoff
formula\cite{ABformula1963}, $I_0 = \pi\Delta G_n/(2e)$, where $\Delta$
is the superconducting (pairing) gap and $G_n$ is the normal state
Josephson junction conductance.

In contrast, a double-well BEC is a closed system with only the total
number of atoms fixed and is not necessarily in the thermodynamic
limit. Deep in the condensed state and for deep wells the coherent
dynamics can be studied via two coupled (Gross-Petaevskii) equations
of motion for two coherent-state amplitudes $\Psi_{L,R}(t)$, with
$|\Psi_{L,R}|^2=N_{L,R}(t)$ giving the number of atoms in the left and
right wells, with only the total number $N=N_L+N_R$ of atoms
conserved. The dynamics is given by Euler-Lagrange equations for the
coherent-state action given by
\begin{equation} \label{eq:action}
S=\int dt\left[\Psi_L^*i\hbar\partial_t\Psi_L +
    \Psi_R^*i\hbar\partial_t\Psi_R - H(\Psi_L,\Psi_R)\right],
\end{equation}
with the Hamiltonian
\begin{eqnarray} 
\label{eq:ham}
 H&=&-J \Psi^*_L\Psi_R  -J  \Psi^*_R\Psi_L + \eps_L|\Psi_L|^2+
  \eps_R|\Psi_R|^2 \\
&&+\frac{g_0}{2N} |\Psi_L|^4 +\frac{g_0}{2N} |\Psi_R|^4 + 
 \frac{g_1}{N} |\Psi_L|^2|\Psi_R|^2,\nonumber
\end{eqnarray}
and parameters $J,\eps_{L,R},g_i$ straightforwardly derivable from a
continuum model of interacting bosons trapped in a double-well
potential\cite{SmerziJEprl1997}. Since the Hamiltonian should scale
linearly with the system size, for convenience we defined the
interaction couplings with explicit factors of $1/N$ so that the
parameters $g_i$ do not scale with the system size.

For $g_i=0$ the dynamics reduces to that of a two-level system (e.g.,
spin in a magnetic field), with oscillations (that we will loosely
call Rabi oscillations) arising from non-eigenstate initial
conditions.  This two level system has a matrix representation
\begin{equation}  \label{eq:matrix}
{\cal H}=\begin{pmatrix}
\eps_L & -J \\
-J& \eps_R
\end{pmatrix}
\end{equation}
with well-known eigenvalues
\begin{equation}
E_{\pm} = \eps \pm \sqrt{J^2+h^2},
\end{equation}
where we defined 
\begin{equation} 
\eps=\oh(\eps_L+\eps_R), \ h=\oh(\eps_L-\eps_R)
\end{equation}
(with the energy difference $h$ not to be confused with the Planck
constant, for which we will use the symbol $2\pi\hbar$ throughout).
Thus any quantity quadratic in $\Psi_L$ and $\Psi_R$ will oscillate
with the Rabi frequency
\begin{equation} 
\label{eq:fre} 
\omega_R = (E_+-E_-)/\hbar = 2 \sqrt{J^2+h^2}/\hbar.
\end{equation}

Although this double-well BEC system was studied in a seminal work by
Smerzi, et al.\cite{SmerziJEprl1997} and a number of works that
followed, from the recent discussion\cite{BEC2009phillips} of the
latest experiment\cite{SteinhauerJEnature2007} it appears that a
number of questions remain unanswered, namely: (i) How is the Rabi
frequency, $\omega_R$, which depends on system-specific quantities
such as $J$, related to the universal Josephson frequency, $\omega_J$ in
(\ref{eq:vf}), which depends only on the applied voltage?  (ii)
Moreover, what role do interactions (clearly neglected in $\omega_R$)
and the thermodynamic limit play in establishing the relation between
the two systems and corresponding frequencies?

A connection between these (otherwise quite distinct) systems only
exists in the specific limit of macroscopic wells,
$L_{L,R}\rightarrow\infty$ (reached for large occupation $N_{L,R}\gg
1$) and a finite barrier thickness $d\ll L$ of the double-well BEC
system. This is necessary in order to approximately model the
thermodynamically large lead reservoirs of a superconducting Josephson
junction. Our key observation is that in this limit the Josephson
coupling, $J$, as defined by \rf{eq:ham} vanishes as $1/L$, i.e.,
vanishes in the large atom occupation number limit, and thus
$\omega_R$ goes over to $\omega_J$, exactly in the thermodynamic
limit, with the identification of $h$ with $eV$.  Moreover, in this
limit the amplitude of oscillations in the number imbalance is always
small regardless of the interaction strength, and thus the
nonlinearity plays negligible role in their dynamics, aside from just
redefining the relationship between $h$ and $V$ to be
\begin{equation} \label{eq:id} 
eV\leftrightarrow h + \left( g_0-g_1\right) \frac{N_L^{(0)} - N_R^{(0)}}{2 N},
\end{equation}
with $N_{L,R}^{(0)}$ the equilibrium number of atom in the two wells.

The vanishing of $J$ for macroscopic leads and fixed barrier width $d$
can be understood on general grounds simply by noting that $J$ is an
interfacial energy per particle, associated with the coupling of the
left and right leads. It is therefore proportional to the surface area
of the barrier $A\sim L^2$ divided by system's volume, i.e., $J\sim
L^2 \ell/L^3\sim \ell/L$, (with $\ell$ the inverse penetration length
scale set by the barrier width and height) and thus indeed vanishes in
the above thermodynamic limit.

To derive these results more explicitly, we estimate the typical size
of $J$ in a macroscopic system by solving the Schr\"odinger equation
of a particle of mass $m_a$ in
 a double-well symmetric potential, $U(z)=U(-z)$, with the
difference between the first excited and ground states by definition
giving $2J$.  This problem is set up in
Ref.~\onlinecite{LandauLifshitz} and consists of constructing wave
functions $\psi_0(z)$ and $\psi_0(-z)$ localized in the left and right
wells, respectively. It is then shown in a straightforward way that
the energy splitting in this double well is given by
\begin{equation} J = -\frac{\hbar^2}{m_a} \psi_0(0) \psi_0'(0).
\end{equation}
To evaluate this expression, it is crucial to distinguish cases of a
smooth and sharp barriers. The case of a smooth barrier is solved in
Ref.~\onlinecite{LandauLifshitz} using the WKB approximation and leads
to the expression for $J$ proportional to the natural frequency of the
oscillations (attempt frequency) in each of the wells as well as the
dimensionless coefficient of penetration through the barrier.  In
contrast, motivated by the connection to the JJ problem we are instead
interested in a sharp barrier (relative to the size $L$ of the wells)
of width $d$, located between points $z=-d/2$ and $z=d/2$ ($z$ is the
axis along the leads and perpendicular to the barrier), where the potential energy exceeds the kinetic
energy by the amount $U_0$. For this setup the
WKB is clearly inapplicable. Under such conditions, the wave function
in the wells is approximately
\begin{equation} 
\psi_0(z) \approx \frac{1}{\sqrt{L}} e^{ik z}, \, z<-\frac d 2,
\end{equation}
where $k$ is the wave vector in the wells, while the wave function
under the barrier is
\begin{equation} 
\psi_0(z) \approx  \frac{1}{\sqrt{L}} e^{-(z+d/2)/\ell}, \, -\frac d 2 \le z \le 0,
\end{equation}
with length $\ell = \hbar/\sqrt{2m_a U_0}$.
This then gives\cite{LLdw}
\begin{equation}
J = \frac{\hbar^2}{m_a \ell L} e^{-d/\ell}.
\label{J}
\end{equation}

Another more direct way of producing \rfs{J} is to note that $J$ is
proportional to the matrix element of the Hamiltonian (e.g., the
kinetic energy) between the left and right wavefunctions, normalized
in each of the wells and penetrating a distance $\ell$ into the
barrier. Clearly then $J\approx\int_{-L}^{L}
dz\frac{\hbar^2}{2m_a}\psi_L^*\psi''_R$ giving result \rf{J}, and as
advertized vanishes for macroscopically large wells.

Consequently, for any realistic energy difference $2h$ between the two
wells, $J$ in (\ref{eq:fre}) can be neglected, reducing the Rabi
oscillation frequency for such large wells to
\begin{equation} 
\omega_R =  2h /\hbar,
\end{equation} 
and allowing the identification of the imbalance $h$ in the
double-well BEC with the chemical potential difference (voltage) $e V$,
\rf{eq:vf}.

We now turn to the analysis of the effects of interactions on our
conclusion above by considering the Hamiltonian (\ref{eq:ham}) with
$g_i \not = 0$.  This is most conveniently done in the (polar)
density-phase representation, by introducing
\begin{equation} 
\label{eq:af1}  
\Psi_L = \sqrt{N_L} e^{i \phi_L t}, \ \Psi_R =\sqrt{N_R} e^{i \phi_R t},
\end{equation}
and
\begin{eqnarray} 
\label{eq:af2}
N_L =\frac N 2 (1 + m), &&  \phi_L = \theta+\phi,\\
N_R =\frac N 2 (1 - m), && \phi_R = \theta-\phi, \nonumber
\end{eqnarray} 
where $-1 \le m \le 1$.  The total number of particles $N$ is
conserved (hence $H$ is independent of $\theta$), and the part of the
Hamiltonian describing the evolution of $m$ becomes
\begin{equation}
 \cH = \frac H N = -J  \sqrt{1-m^2}\cos(2\phi)
+  h  m + \frac{\lambda}{2} m^2,
\end{equation}
where $\lambda=(g_0-g_1)/2$.  The equations of motion are thus
\begin{eqnarray} 
  \label{eq:eom} 
  \hbar \dot \phi &=& - \frac{\partial
    \cH}{\partial m},\nonumber\\
&=& - h - \lambda  m - J\frac{m}{\sqrt{1-m^2}} \cos(2\phi)\equiv\frac{\delta\mu}{2},\hspace{1cm}\\
  \hbar \dot m &=&
  \frac{\partial \cH}{\partial \phi}=  2 J \sqrt{1-m^2}
  \sin(2\phi), \label{eq:mdot}
\end{eqnarray}
with \rf{eq:eom} defining the effective chemical potential difference
$\delta\mu = \mu_L-\mu_R$. Although it is not fixed in this canonical
ensemble, in the thermodynamic limit of interest its constant (in
time) part can be meaningfully associated with the grand-canonical
chemical potential different of the JJ system.  Above equations
provide a complete solution to the problem of two couple interacting
condensates and have been analyzed in
Ref.~\onlinecite{SmerziJEprl1997}.  Although it is possible to solve
these equations analytically, the general solution is not very
informative.

Here we focus on the physical limit of large reservoirs discussed
above, in which the coupling $J$ is vanishingly small. In this limit
we solve Eqs.~\rf{eq:eom} perturbatively in powers of $J/h\ll 1$.  At
zeroth order in $J/h$,
\begin{eqnarray}
\hbar \dot{\phi}_0 &=& - h - \lambda  m_0,\\
\dot{m}_0 &=& 0,
\end{eqnarray}
leading to the solution
\begin{eqnarray}
\label{zeroOrder}
\phi_0(t) &=& - \frac{h + \lambda  m_0}{\hbar} t \equiv -\h t,\\
m_0 &=& \text{const.}
\end{eqnarray}
Identifying atom current $I$ with $(\dot N_R - \dot N_L)/2=-N\dot m/2$, we
obtain, with the help of \rfs{eq:mdot},
\begin{equation} 
\label{eq:I} 
I  = \frac{ J N\sqrt{1-m_0^2}}{\hbar}
\sin\left(\frac{2\left( h+\lambda m_0 \right) t}{\hbar} \right), \
\end{equation}
that is clearly equivalent to \rf{eq:vf}, with the identification
\rf{eq:id} and $I_0=JN \sqrt{1-m_0^2}/\hbar$.  

From this last identification of $I_0$ we can further note that the
result \rfs{J} is compatible with the Ambegaokar-Baratoff expression
for the critical current in a Josephson junction, namely, that
\begin{equation} I_0 \sim \frac{N}{L\ell} = \frac{\rho L^2}{\ell},
\end{equation} 
with $\rho = N/L^3$ the atom density. From this we observe that the
critical current of the double-well BEC scales as the area $L^2$ of
the ``junction'' (double-well barrier), just like the
Ambegaokar-Baratoff expression for the superconducting 
JJ, $I^{AB}_0 =
\pi\Delta G_n/(2e)$, where the junction area enters through the number
of conduction channels in $G_n$. This further supports our finding
that $J$, as defined in \rf{eq:ham} vanishes with the inverse length of
the well ``leads''.
 
Let us see whether any realistic Josephson junctions indeed obey the
condition $J \ll h$.  For a typical Josephson junction, $d \sim 100$nm
\cite{Likharev}. $\ell$ must itself be of the order of $d$, otherwise
the coupling $J$ will be even further exponentially suppressed. Taking
$L \gtrsim 100$nm and $h \sim 1$eV, we find
\begin{equation}
\frac{J}{h} = \frac{\hbar^2}{m_a \ell L h} \lesssim 10^{-5},
\end{equation}
where an electron mass was used for $m_a$. Roughly, the above factor
of $10^{-5}$ arises due to two factors of $10^{3}$ of $\ell$ and $L$
(large characteristic length scales) relative to the Bohr radius that
corresponds to an eV energy scale. Thus indeed in a realistic
Josephson junction $J$ is always much smaller than $h$, well
justifying above approximation, even if the system were to be
closed. Of course, as mentioned in the introduction, a conventional
superconducting Josephson junction is an open system, that is a part
of a macroscopic circuit and is therefore effectively characterized by
an infinite $L$. Furthermore, with the circuit driven by a fixed
voltage source the Josephson frequency expression $\omega_J$,
\rf{eq:vf} is effectively exact; any inaccuracy in Josephson
voltage-frequency relation quoted above is associated with the
uncertainty of the current knowledge of Planck's constant (one part in
$10^7$)\cite{SamBenz}.

In contrast a typical double-well BEC trap potential is expected 
to be roughly characterized by a
single (the same order of magnitude) length and energy scale, with
$\ell\sim L$ and $U_0\sim h$, leading to $J$ and $h$, that are
comparable and both a tiny fraction of an electron volt. Consequently,
we expect such a system to display a significant and tunable deviation
from the Josephson frequency, \rfs{zeroOrder}, obtained by neglecting
$J/\h$ corrections. 

Even though as discussed above in a double-well BEC system we
generically expect $J\sim h$, we observe that the 
experimentally studied double-well BEC\cite{SteinhauerJEnature2007} is
characterized by $J/\hbar\approx 15$ sec$^{-1}$, and $1400\lesssim
h/\hbar\lesssim 5700$ sec$^{-1}$ (here and throughout $h$ denotes the
chemical potential imbalance energy, not the Planck's constant
$2\pi\hbar$), and thus corresponds to $0.003\lesssim J/h\lesssim
0.011$.\cite{emailJeff} This small value of $J/h$ characterizing these
experiments explains why the measured oscillation frequency
($\omega_{acJ}$) as a function of the chemical potential difference
($h$) is observed to be linear in
Ref.~\onlinecite{SteinhauerJEnature2007}. In order to detect a
deviation from this linear behavior these measurements need to be
extended down to $h/\hbar\lesssim 15$ sec$^{-1}$, or done on a system
in which the value of $J$ is increased by e.g., making wells smaller.

With this in mind, it is useful to compute the lowest order
correction.  This can be straightforwardly done by evaluating the
solutions $\phi(t), m(t)$ to \rfs{eq:eom} systematically to $n$th
order in $J/h$ by iterating the equations, with the $n$th-order
solution $\phi_{n}(t), m_{n}(t)$ on the left-hand side and
approximating the right-hand side by the $n-1$st-order solution,
$\phi_{n-1}(t),m_{n-1}(t)$.

To first-order the equations become:
\begin{eqnarray}
\dot{\phi}_1&=&-\h - J\frac{m_0}{\sqrt{1-m_0^2}}\cos(2\phi_0),\\
\dot{m}_1&=&  2J\sqrt{1-m_0^2}\sin(2\phi_0),
\end{eqnarray}
leading to the solution
\begin{eqnarray}
\phi_1(t) &=& - \h t -
\frac{J}{2\h}\frac{m_0}{\sqrt{1-m_0^2}}\sin(2\h t),\\
m_1(t) &=& m_0 +
\frac{J}{\h}\sqrt{1-m_0^2}\cos(2\h t).
\end{eqnarray}

The solution to second-order, $(J/\h)^2$ is obtained by using above
expressions, $\phi_1(t), m_1(t)$ in the right-hand side of the exact
equations \rf{eq:eom}:
\begin{eqnarray}
\dot{\phi}_2&=&-\h - J\frac{m_1}{\sqrt{1-m_1^2}}\cos(2\phi_1),\\
\dot{m}_2&=&  2J\sqrt{1-m_1^2}\sin(2\phi_1).
\end{eqnarray}
Focusing on $\phi(t)$ and integrating the first equation we find the
time-independent part of $\dot\phi$ 
\begin{widetext}
\begin{eqnarray}
  \dot{\phi}_2&=&-\h -
  J\frac{m_0+\delta m(t)}{\sqrt{1-(m_0+\delta m(t))^2}}
  \cos\left[2\h t+\frac{J}{\h}\frac{m_0}{\sqrt{1-m_0^2}}
    \sin(2\h t)\right],\\
  &\approx&-\h
  -\frac{J m_0}{\sqrt{1-m^2_0}}\cos(2\h t)
  +\frac{J^2}{\h}\frac{m^2_0}{1-m_0^2}\sin^2(2\h t)
  -\frac{J^2}{\h}\frac{1}{1-m_0^2}\cos^2(2\h t),\\
  &\approx&-\h
  +\frac{J^2}{2\h}\frac{m^2_0}{1-m_0^2}
  -\frac{J^2}{2\h}\frac{n^2}{1-m_0^2}+\ldots,\\
  &\approx&-\h-\frac{J^2}{2\h}\approx -\sqrt{\h^2 + J^2}, 
\end{eqnarray}
\end{widetext}
which by definition \rf{eq:eom} is the effective chemical potential
difference. In above we have neglected the higher order hamonics that
are also always generated at nonzero $J/\h$. This then gives
\begin{eqnarray}
\phi(t)&\approx&-t\sqrt{\h^2 + J^2},\\
\dot{m}&\approx& -2J\sqrt{1-m_0^2}\sin\left[2t\sqrt{\h^2 + J^2}\right].
\end{eqnarray}
We therefore obtain the (fundamental) ac Josephson frequency of
current oscillations (defined by $\dot{m}$) to be given by
\begin{eqnarray}
\label{eq:omegaAC}
\omega_{acJ}&\approx&\h+\frac{J^2}{2\h}
\approx\sqrt{\h^2 + J^2}, 
\end{eqnarray}
with $\h = h + \lambda m_0$.  

Thus, as advertized, for macroscopically occupied wells (the only
limit in which a reasonable connection to a Josephson junction can be
made), such that $J/\h\ll 1$, an imbalanced double-well BEC
system is indeed a good model for the ac Josephson effect, exhibiting
current oscillations with frequency that is nearly independent of $J$
and grows linearly with imbalance $\h=h+\lambda m_0$. However,
for a smaller double-well BEC (more typical experimentally) the
fundamental oscillation frequency is expected to exhibit
$(J/\h)^2$ deviations from the linear dependence on $\h$
of the form given in \rf{eq:omegaAC}.

We thank S. Benz, K. Lehnert, W. Phillips, and J. Steinhauer for
discussions, appreciate the hospitality of the Kavli Institute for
Theoretical Physics in China during the "Condensed Matter Physics of
Cold Atoms" workshop, when part of this work was performed, and
acknowledge financial support by the NSF grants DMR-0321848 (L.R.),
DMR-0449521 (V.G.), and PHY-0904017 (V.G.).

\end{document}